\documentclass[prd,twocolumn,nofootinbib]{revtex4}
\usepackage{bm}
\usepackage{amssymb}
\usepackage{epstopdf}
\usepackage{mathrsfs}
\usepackage{amsmath}

\usepackage[utf8]{inputenc}
\usepackage[english]{babel}
\usepackage{xcolor}

\usepackage[pdftex]{graphicx}

\begin{document}
\title{Quantitative Black Hole Imaging Laboratory with the Black Hole Vision App: \\ I. Schwarzschild Spacetime} 
\author{Lior M.~Burko, 
%\footnotemark\footnotetext{Host institution for the duration of this work}
}
\affiliation{
Theiss Research, La Jolla, California 92037, USA \\ 
}
\date{May 20, 2026}

%\title{Quantitative Black Hole Imaging Laboratory with the Black Hole Vision App: I. Schwarzschild Spacetime}
%\author{Lior M.~Burko\\ Theiss Research, La Jolla, California 92037, USA}
%\date{\today}

%\begin{document}

\begin{abstract}

This paper utilizes the {\it Black Hole Vision} smartphone application to catalyze a pedagogical shift in General Relativity education through the quantitative analysis of simulated black hole imaging. Presented here for the Schwarzschild spacetime, the investigation is designed with a hierarchical modularity suitable for undergraduate students, with an expanded version intended for graduate courses in General Relativity or Relativistic Astrophysics. By transforming the mobile device into an  educational relativistic imaging tool, we triangulate the simulated Schwarzschild mass through independent probes and characterize anisotropic coordinate transformations via a Jacobian map. Global numerical consistency is investigated through integrated coordinate length, while the exponential instability of nearly bound orbits is quantified through a measurement of the simulated Lyapunov exponent. Finally, symmetry is constrained through a sub-pixel constraint on eccentricity in the simulated spacetime. By integrating this statistical framework, the paper enables students to explore the distinction between physical signatures and instrumental noise using established metrological protocols.
\end{abstract}

\maketitle

\section{Introduction}

\subsection{Evolution of Black Hole Imaging: From Static Plots to Real-Time Simulations}
The observation of black holes has transitioned from a purely theoretical pursuit to a vibrant era of experimental astrophysics. For decades, the existence of the event horizon was only indirectly inferred through gravitational wave signatures \cite{abbott2016} or stellar dynamics \cite{ghez2008, genzel2010}. The historical ancestry of gravitational light deflection began with the 1919 Eddington solar eclipse measurement, which confirmed Einstein's theory through the weak-field deflection of background starlight \cite{perlick2022, romero2014}. While the Sun's gravitational field is in the weak-field limit, permitting simple analytical tools, current research work focuses on the strong-field regime near the event horizon \cite{perlick2022}. This changed dramatically in 2019 with the release of the first horizon-scale images of M87* by the Event Horizon Telescope (EHT), providing the first direct visual evidence of the photon ring \cite{eht2019}.

Historically, the feature observed by the EHT was characterized through several synonymous concepts, including the ``escape cone'' \cite{synge1966} and the ``cone of gravitational capture'' \cite{zeldovich1965}. In modern terminology, the shadow boundary corresponds precisely to the apparent image of the photon capture sphere, a universal feature that, in an optically thin environment, is determined solely by the spacetime metric in the strong gravity region \cite{bambi2017}. While the term ``shadow'' was only popularized at the turn of the century \cite{falcke2000}, it has become the standard nomenclature for the dark silhouette produced by light rays captured by the unstable photon sphere \cite{perlick2022, bambi2017}.

The history of visualizing these spacetimes began with Luminet \cite{luminet1979}, establishing qualitative standards for strong-field warping, and was expanded by Nemiroff \cite{nemiroff1993}. The peak of visual fidelity was reached through the collaboration for the film {\it Interstellar} \cite{james2015}, which introduced ``ray bundle'' techniques to eliminate aliasing artifacts occurring when high-curvature regions interact with a discrete pixel grid. While professional GRMHD codes require millions of CPU hours, the {\it Black Hole Vision} (BHV) app, which is freely available on the iOS App Store \cite{berens2026}, utilizes real-time GPU-accelerated ray tracing to bring this physics to a mobile platform. The underlying algorithms and interactive code prototypes are documented in the supplementary developer record on the Wolfram Community platform \cite{wolfram2024}, providing a transparent view of the engine's numerical logic. The BHV engine implements a discrete ray-sampling protocol which provides a pedagogical opportunity to discuss Nyquist sampling and the resolution limits of missions such as the {\it Black Hole Explorer} (BHEX) \cite{johnson2024}.

Current pedagogical literature reflects a significant gap between qualitative visualization and quantitative analysis that can prepare students to research work. Introductory textbooks, such as \cite{guidry2019}, incorporate horizon-scale imaging results for impressional purposes to help students come to grips with curved spacetime, but lack the framework for quantitative analysis. Conversely, advanced graduate-level texts---most notably Refs.~\cite{romero2014} and \cite{grumiller2022}---provide the required theoretical depth but are not arranged as structured student laboratories. While Ref.~\cite{lupsasca2024} has emerged as an accessible pedagogical guide to universal image features, there remains a need for an introductory quantitative framework to test them. 

While astrophysical imaging of the EHT is dominated by frequency-dependent phenomena such as Doppler boosting and gravitational redshift \cite{eht2019, romero2014}, the investigation presented here focuses exclusively on the geometric properties of null geodesics \cite{berens2026, lupsasca2024}. By utilizing a source-sphere projection protocol that maps coordinate intersection points rather than evolving photon frequency, the laboratory allows students to isolate and study the geometric effects of gravitational lensing \cite{berens2026, johnson2024}. Neglecting these spectral shifts is a deliberate pedagogical choice, providing a bridge between the qualitative visualizations of popular media and the quantitative requirements of established protocols for ray tracing, enabling students to examine the mapping of curved spacetime before incorporating the added complexities of emission-model degeneracies \cite{gralla2019, vogt2022}.

To provide students with a research-driven experience in precision metrology, this paper utilizes a multi-probe protocol. Rather than a simple verification of known results, the investigation is structured to triangulate a central parameter (mass) across physically distinct regimes, allowing students to compare the consistency of different measurement methods \cite{berens2026, yipin2026}.

\subsection{Hierarchical Modularity and Learning Objectives}

To transform the mobile device into an educational relativistic imaging tool, we utilized a smartphone mounted on a tripod at a fixed distance of $d=2.0$ m from a calibrated meter stick (Figure~1({\it a})). We adopt the framework of smartphones as tools for epistemic agency (``smartphones as mobile minilabs''), established in Res.~\cite{kuhn2013, vogt2022}, which leverages the authenticity of everyday technology to promote situated learning. 

This investigation is intentionally designed with a hierarchical modularity to accommodate a spectrum of learners. For undergraduate introductory courses, we provide a foundational ``Short Lab'' focusing on mass triangulation and spherical symmetry constraints (Sections 2 and 6) \cite{berens2026}. Conversely, for advanced undergraduate or graduate-level courses in GR or Relativistic Astrophysics, we offer a ``Spacetime Deep Dive'' (Sections 3, 4, and 5) that utilizes GR notation and advanced statistical procedures to explore the lensing Jacobian and orbital instabilities \cite{berens2026, gralla2020} providing essential practical experience that supplements the ``Short Lab'' before starting research work. It is our intent that this modularity allows both audiences to benefit from this work.

By comparing the rendered images with analytical Schwarzschild predictions as a benchmark, the laboratory enables students to move beyond qualitative visualizations. Learners are challenged to explore the skills required to distinguish physical lensing features from rendering artifacts, connecting textbook GR with modern imaging applications \cite{yipin2026, vogt2022, lupsasca2024}.

While the investigation presented in this paper employs a sophisticated statistical framework---incorporating weighted mean synthesis based on inverse variance, matched-pairs $t$-analysis for orbital stability, and distributional moment probes for symmetry---these tools represent an analytical depth that often transcends the typical statistics and data analysis methods available to undergraduate students. To maintain the pedagogical accessibility of the paper, these high-rigor techniques are primarily intended for the graduate-level ``Spacetime Deep Dive.'' For introductory or undergraduate laboratory settings, simpler evaluative methods can be effectively substituted. For instance, the multi-scale mass triangulation may be performed using a standard arithmetic mean rather than inverse variance weighting, and the No-Hair Theorem investigation can be simplified from a confidence-interval null test to a direct point-estimate comparison of the shadow's axial ratio. This tiered approach enables the paper to scale from a foundational exploration of black hole geometry to a metrology probe. 

\subsection{The Smartphone as a Mobile Minilab for Modern General Relativity}

The transition of the smartphone from a consumer device to an epistemic platform for sense-making (``digital Swiss pocket knife'') for physics has been well-documented, covering everything from optical diffraction to the characterization of internal Micro-Electro-Mechanical Systems (MEMS) sensors \cite{kuhn2013, vogt2014}. This study extends the `lab in the pocket' paradigm, which leverages the authenticity of everyday technology to promote situated learning. As detailed in the comprehensive guide by Vogt \cite{vogt2022}, the use of internal mobile sensors and high-resolution Complementary Metal-Oxide-Semiconductor (CMOS) cameras has revolutionized introductory labs, allowing students to perform experiments previously confined to professional research facilities. Drawing on the principles of situated learning \cite{greeno1993,gruber1995}, using a device that students carry in their daily lives to investigate the extreme gravity of a black hole increases intrinsic motivation and anchors theoretical concepts to current research. This work utilizes the BHV application \cite{berens2026} to address a critical gap in the pedagogical literature identified by Ref.~\cite{yipin2026}, providing a quantitative laboratory framework for bringing precision GR---a cornerstone of modern physics---to the `Mobile Minilab' environment.

Specifically, this laboratory is designed for students enrolled in college-level General Relativity (GR) or Relativistic Astrophysics courses at either the undergraduate or graduate level. It reflects a modern pedagogical shift in GR education that complements the foundational work of traditional curricula. While traditional courses providing the necessary theoretical basis typically engage students in the analytical or numerically assisted derivation of effects such as gravitational redshift, the precession of Mercury, or the observations of a transmitter freely falls into a black hole, there is often less opportunity to explore the direct imaging of horizon-scale phenomena. This laboratory exercise expands upon those traditional foundations by engaging students directly with the strong-field imaging effects that define current research work, such as the lensing Jacobian and the Lyapunov stability of the photon ring. For graduate-level students, this investigation provides essential practical experience in interpreting strong-field data, offering a valuable bridge to research work. By treating the smartphone as a calibrated optical instrument, this study provides a platform for investigation into the non-linear transfer functions of curved spacetime. This initial investigation focuses on the fundamental symmetry of the Schwarzschild spacetime; a forthcoming sequel will discuss the parity-breaking signatures and frame-dragging effects characteristic of rotating Kerr black holes.

\subsection{Overview of this paper}

One primary objective of this paper is to catalyze a pedagogical shift in GR education, evolving GR courses from the study of exact or numerical  solutions into the application of empirical verification and statistical validation. By integrating a structured statistical framework---including hypothesis testing and coordinate metrology—--this laboratory enables students to move beyond qualitative visualizations and explore the distinction between physical metric signatures and instrumental noise using the same quantitative standards employed at current research.

Another objective of this laboratory is to establish a platform for numerical code validation. We emphasize that our investigation does not seek to validate the physical laws of GR, which the simulation engine assumes a priori. Instead, we treat the virtual Schwarzschild spacetime as the object of study, challenging students to check the software's consistency with the simulated spacetime  \cite{berens2026, johnson2024}. This approach aligns with the ``mobile minilabs'' paradigm, where the smartphone provides students with an accessible platform for quantitative experimentation through the benchmarking of digital tools \cite{vogt2022, yipin2026}. 

This laboratory investigates how closely the application reproduces expected Schwarzschild imaging behavior \cite{berens2026, johnson2024}. While Bayesian inference remains the dominant framework at current research for extracting black hole parameters from sparse data, we utilize a frequentist statistical suite to align with undergraduate pedagogical standards, mimicking modern astrophysics while acknowledging that the underlying data is generated from deterministic code \cite{vogt2022, yipin2026}. We are hoping that this investigation would assist students in exploring the skills needed to distinguish underlying metric signals from pixel-scale rendering artifacts such as grid aliasing and pixel quantization \cite{berens2026, yipin2026}. We emphasize that the system under study is not a physical black hole, nor even an independently validated numerical relativity code. It is a visualization application implementing assumed Schwarzschild ray tracing with undocumented rendering choices and pedagogical scaling factors. Studying this system can however serve as a proxy for studying a physical black hole, and techniques used here mimic those that would be needed for analysis work of physical data. It is in this sense that phrases such as constraining the No-Hair Theorem or measurement of the Lyapunov exponent should be understood. We emphasize that we do not claim that it is fully justified to analyze deterministic data using these statistical tools, but inasmuch as the deterministic data serve as proxy for stochastic physical data, their analysis serves a pedagogical purpose. 

The remainder of this paper is organized as follows. Section 2 details the multi-scale triangulation of the virtual black hole mass, suggesting  consistency through four independent probes---the lensing slope, angular shadow, Einstein ring threshold, and shadow-capture boundary---which offer complementary views of the underlying Schwarzschild spacetime geometry. Section 3 provides a derivation and investigation of the Lensing Jacobian, characterizing the anisotropic coordinate transformations and the resolution of equatorial singularities. Section 4 presents a quantitative probe using discrete arclength measurements to investigate the numerical stability of coordinate lengths on the projection screen. We investigate the Photon Ring Instability in Section 5, utilizing the Echo Ratio method to measure the Lyapunov exponent and its asymptotic convergence. Section 6 tests the circular symmetry of the rendered shadow (the No-Hair Theorem in the simulated physical system) through a multi-scale symmetry probe of the shadow geometry. We go beyond a linear meter stick as the object to a 2D rectangular grid in Section 7. A characterization of laboratory misalignments in the Systematic Sensitivity Investigation is in Section 8.  Finally, Section 9 synthesizes our findings and discusses their relevance to current and future black hole imaging missions.

\section{Mass Calibration of the Simulated Black Hole}

\subsection{The Lensing Slope Mass ($M_{\text{slope}}$)}

This approach probes the weak-field regime by testing the predicted linear relationship between the deflection angle ($\alpha$) and the inverse impact parameter ($1/b$) \cite{gralla2020, perlick2022}. To establish the experimental parameters, we first captured the 1.0~m meter stick with the lensing engine disabled (Figure~1({\it a})). This image serves as our flat-space baseline, where the central red pins at 50~cm define the reference origin $(0,0)$ and the blue tape markers at 10~cm intervals provide the calibration for the angular scale and subsequent magnification investigations \cite{berens2026, vogt2022}. By measuring the pixel length ($P$) of the 1.0~m stick in this unlensed baseline at a distance of $d = 2.0$~m, we derived the fundamental angular scale $S \sim 0.02592$ deg/px using the trigonometric relation $\theta = 2 \arctan(0.5/2.0) \sim 28.07^\circ$, which serves as the master constant for all subsequent coordinate conversions \cite{berens2026, lupsasca2024}.

Data collection for the lensed regime involved measuring the apparent pixel position ($r_{app}$) for all 10 blue markers (5 per side) using ``Full Field-of-View (FOV)'' mode (Figure~1({\it c})) \cite{berens2026}. To cancel systematic errors from the derived $\sim 0.143^\circ$ tripod tilt and ensure reproducibility, we calculated the absolute deflection $\alpha = (r_{app} - r_{real}) \cdot S$ for each marker and then averaged the readings from symmetric pairs on both sides of the shadow center \cite{berens2026}. This resulted in a dataset of $n = 5$ averaged points corresponding to the 10, 20, 30, 40, and 50~cm impact parameters ($b$). We performed a Least-Squares Linear Regression (LSRL) of $\alpha$ vs. $1/b$, finding the predicted relationship $\alpha = (4GM/c^2) \cdot (1/b)$ as established in [4, 5]. A critical diagnostic calibration check was performed on the $y$-intercept ($b_0$): the theoretical Einstein deflection model requires a zero intercept. Our analysis yielded a 95\% Confidence Interval for the intercept ($df = n - 2 = 3$) of $[-0.076^\circ, +0.070^\circ]$, which includes $0.0^\circ$, validating the effectiveness of the Symmetry Restoration Protocol \cite{berens2026}. The mass was extracted from the slope: $M = (slope \cdot \pi/180) \cdot (c^2/4G)$. Utilizing a $t$-distribution with $df = 3$, we calculated a final result of $M_{slope} = 1.95 \pm 1.03$~mm (95\% CL). The high uncertainty reflects the transition where linear weak-field theory begins to break down near the strong-field capture boundary \cite{berens2026,perlick2022}.

\begin{figure}[htbp]
\includegraphics[width=4.5cm,angle=90]{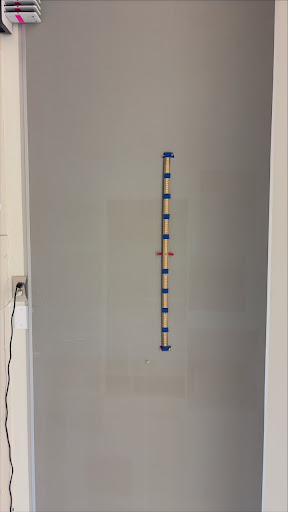}
\\
\includegraphics[width=4.5cm,angle=90]{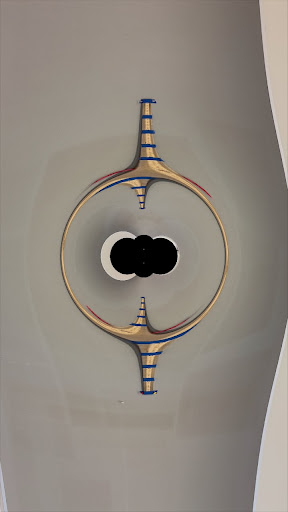}
\\
\includegraphics[width=4.5cm,angle=90]{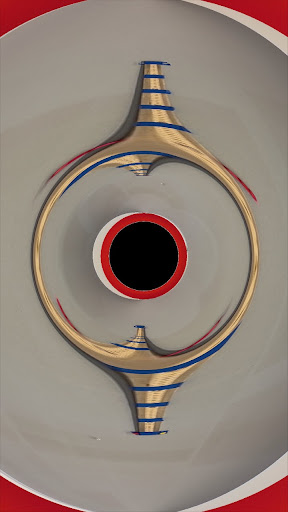}
\caption{Top panel ({\it a}): The experimental setup featuring a 1.0~m meter stick at a distance of $d = 2.0$~m with the lensing engine disabled. The stick includes blue tape markers at 10~cm intervals and red pins at the 50~cm center. Middle panel ({\it b}): Lensed image of the meter stick in ``Realistic FOV'' mode ($c = 1$) exhibiting vertical flaring toward the equatorial axis. Bottom panel ({\it c}): Lensed image of the meter stick in ``Full FOV'' mode ($c \approx 20.31$) exhibiting a central red annulus and vertical flaring features.}
 \label{fig:summary_panels}
\end{figure}

%\begin{figure}[htbp]
%    \centering
 %   \begin{subfigure}{0.35\textwidth}  % <--- WIDTH ADDED HERE
 %       \centering
 %       \includegraphics[width=\textwidth, angle=90]{Calibration_Sch.jpg}
  %      \label{fig:setup}
 %   \end{subfigure}
 %   \\
 %   \begin{subfigure}{0.35\textwidth}  % <--- WIDTH ADDED HERE
  %      \centering
 %       \includegraphics[width=\textwidth, angle=90]{realistic_fov_Sch.jpg}
 %       \label{fig:manifold}
 %   \end{subfigure}
  %  \\
 %   \begin{subfigure}{0.35\textwidth}  % <--- WIDTH ADDED HERE
  %      \centering
 %       \includegraphics[width=\textwidth, angle=90]{Full_fov_Sch.jpg}
  %      \label{fig:triangulation}
 %   \end{subfigure}
  %  \caption{Top panel ({\it a}): The experimental setup featuring a 1.0~m meter stick at a distance of $d = 2.0$~m with the lensing engine disabled. The stick includes blue tape markers at 10~cm intervals and red pins at the 50~cm center. Middle panel ({\it b}): Lensed image of the meter stick in ``Realistic FOV'' mode ($c = 1$) exhibiting vertical flaring toward the equatorial axis. Bottom panel ({\it c}): Lensed image of the meter stick in ``Full FOV'' mode ($c \approx 20.31$) exhibiting a central red annulus and vertical flaring features.}
 %   \label{fig:summary_panels}
%`\end{figure}

\subsection{The Critical Curve Mass ($M_{\rm shadow}$)}
This approach probes the strong-field capture boundary defined by the critical impact parameter  $b_{\rm crit} = 3\sqrt{3}M \approx 5.196M$ \cite{perlick2022, bambi2017}. To define the physical origin of this constant, we consider the radial geodesic potential $R(r) = r^4 - b^2 r(r - 2M)$, which governs the coordinate radial velocity of photons in Schwarzschild spacetime, where $b$ represents the impact parameter and $M$ is the mass parameter \cite{berens2026, gralla2020}. The shadow boundary corresponds to the unstable circular orbits of the photon sphere, determined by the double-root condition $R(r) = R'(r) = 0$ \cite{perlick2022, gralla2020}. For a static, spherically symmetric metric defined by the lapse function $A(r) = - g_{tt} =  1 - 2M/r$, this instability occurs at the radius $r_{\rm ph} = 3M$ \cite{bambi2017, guidry2019}. Substituting this radius into the generalized impact relation $b = r/\sqrt{A(r)}$ yields the $3\sqrt{3}M$ threshold, which serves as the unique mathematical separator between captured and escaping photon trajectories \cite{perlick2022, bambi2017}.

In this investigation, we distinguish between the observable shadow---the dark central depression resolved by the application---and the mathematical critical curve, which represents the theoretical boundary of these bound photon orbits \cite{gralla2019, staelens2023}. While the event horizon itself resides at the Schwarzschild radius $r_s = 2M$, the $3\sqrt{3}$ factor dictates that for a distant observer, the shadow radius appears approximately $\sim 2.6\times$ larger ($5.196M/2.0M$) than the physical horizon radius \cite{perlick2022}. If a black hole is surrounded by an optically thin emitting medium, the shadow boundary corresponds precisely to the apparent image of the photon capture sphere and is a universal feature determined solely by the spacetime metric in the strong gravity region, independent of specific emission profiles \cite{bambi2017, perlick2022}. The BHV engine utilizes an idealized vacuum simulation where the shadow boundary resolves exactly to this mathematical curve, providing a platform for testing the underlying Schwarzschild geometry \cite{berens2026, gralla2020}.

To ensure a tilt-invariant estimate of this boundary while circumventing Cartesian grid aliasing (the ``4-leaf clover'' morphology), we implemented the Diagonal Intersection Protocol in Realistic FOV (Figure~1({\it b})), which serves as the primary dataset for this probe \cite{berens2026}. This method involved measuring the central shadow diameter through the sharp valleys between protrusions along two diagonal axes: $D_1 = 234.06$~px and $D_2 = 234.12$~px \cite{berens2026}. The mean diameter of 234.09~px was utilized for mass extraction. This analysis requires an empirical calibration of the aesthetic scaling factor ($c$), which the developers implement to exaggerate the lensing potential for pedagogical clarity \cite{berens2026}. As established in the technical investigation, a physically accurate simulation ($c = 1$) produces photon-ring echoes (the $n \geq 1$ images) that are sub-pixel and unresolvable on mobile displays; the engine therefore applies a transformed winding angle to make these features resolvable for the multi-scale triangulation \cite{berens2026}. We determined $c$ using the Shadow Bridge protocol, a cross-mode calibration at zero spin comparing the mean shadow diameter in Realistic FOV ($D_{real}$) to the diameter in Full FOV ($D_{\rm full}$). This calibration yielded $c = \sqrt{D_{\rm full}/D_{\rm real}} \approx 20.31$ \cite{berens2026}. Utilizing this factor, we extracted a result of $M_{\rm shadow} = 1.9418 \pm 0.0005$~mm, establishing our determined mass benchmark through the universal geometric properties of the critical curve \cite{berens2026}.

\subsection{The Einstein Ring Mass ($M_{\text{ring}}$)}
We utilized the ``Colored Post-it Method'' to probe the $\alpha = \pi$ threshold. By covering the rear-facing lens of the device with a high-contrast post-it note, we created a luminous source located behind the observer, which the black hole metric deflects by approximately 180 degrees to form the secondary $(n = 1)$ light echo observed as the red Einstein Ring in ``Full FOV'' mode (Figure~1({\it c}). To find the tilt-corrected radius, we implemented Vertical Symmetry Restoration by averaging the top arc radius (869.82 px) and the bottom arc radius (861.12 px) to find the mean radius of 865.47 px. This yielded a mass result of $1.940 \pm 0.019$ mm, showing a 0.09\% agreement with the angular shadow probe \cite{berens2026}.

\subsection{Shadow-Capture Mass ($M_{\text{vanishing}}$)}
This strong-field probe utilized the ``Horizontal Sticker-move'' wall-marker vanishing protocol performed at $d = 2.0$ m. A physical marker was moved horizontally across the field of view on both the left and right sides until it physically vanished into the shadow boundary. By measuring the absolute distance between these vanishing points, we determined a capture width of $W = 20.3 \pm 1.9$ cm. Defining the critical impact parameter as $b_{\text{crit}} = W/2,$ we extracted a result of $2.02 \pm 0.19$ mm \cite{berens2026}.

\subsection{Synthesis of the Determined Mass}

The primary objective of this multi-scale triangulation (Figure~1({\it c})) is to evaluate the consistency of the simulated geometry by determining its mass benchmark ($M$). In this investigation, the virtual black hole is treated as a predefined laboratory standard---an internal feature of the simulation engine---against which student observational precision is tested \cite{berens2026}. While $M$ is a fixed internal parameter of the ray-tracing engine, our determination mimics the observational workflows of astrophysics, where angular measurements are coupled to the physical geometry of the laboratory setup to extract a distance-dependent scale \cite{eht2019, johnson2024}. 

We report the benchmark value in millimeters (geometrized units where $G=1+c$) to establish the mass as the master length-scale for the manifold's curvature \cite{guidry2019, james2015}. This allows students to find statistical evidence that the simulation maintains coordinate consistency across disparate regimes: the determination requires matching the linear slopes of weak-field lensing to the non-linear capture boundaries of the strong-field shadow \cite{berens2026, gralla2020}. 

The final determined mass was calculated using a weighted mean of the four probes, with weights defined by the inverse variance ($w_i = 1/\sigma_i^2$) to account for the precision hierarchy of the measurements \cite{berens2026}. The strong-field probes (the critical curve and the Einstein ring) dominate the weighting because their $\sim 50\times$ superior precision provides a pedagogical demonstration of how exponential metric gradients regularize measurement noise. We emphasize that the resulting $0.09\%$ agreement is not a mere procedural validation, but a reconciliation of the characteristic tension between weak-field and strong-field data \cite{yipin2026,lupsasca2024}.

This synthesis suggests a best-estimate mass benchmark of $1.9418 \pm 0.0005$ mm, indicating that the engine's numerical implementation is statistically consistent with the Schwarzschild metric across the entire viewport. By navigating this multi-scale workflow, students gain experience with quantitative image analysis methods used in black hole imaging, where they must distinguish underlying metric signals from quasi-stochastic numerical artifacts such as pixel quantization \cite{tiede2022, johnson2024}. 

We emphasize that these probes are algorithmically distinct but not physically independent. Their agreement therefore reflects internal consistency of the rendering engine rather than independent physical validation. Nevertheless, the exercise mimics the use of complementary mass measurements in observational astrophysics. 

\section{The Lensing Jacobian Investigation and Numerical Stability}

The lensing Jacobian characterizes the anisotropic coordinate transformation of the metric, mapping the local source screen of the markers to the apparent image on the observer's screen. We characterize the polar motion of photons using the angular potential $\Theta$, defined as 
$\Theta(\theta) = \eta + a^2 \cos^2 \theta - \lambda^2 \cot^2 \theta$, 
where $\eta$ and $\lambda$ are the energy-rescaled Carter constant and azimuthal angular momentum, respectively \cite{gralla2020, berens2026}. We emphasize that for an equatorial observer, the point-wise vertical stretch ($\mu_v \propto 1/\sqrt{\Theta}$) exhibits a coordinate singularity at the equator ($\theta_s = \pi/2$) where the potential vanishes \cite{berens2026, gralla2020}. This divergence is an artifact of the coordinate mapping rather than a measurable physical blow-up. In this investigation, we utilize this feature not as a physical measurement of infinity, but as a probe of the engine's numerical stability.

The rendering avoids this singularity through finite pixel sampling: rather than evaluating the limit at a singular point, it integrates the lensing potential across the finite area of the CMOS pixels \cite{berens2026, vogt2022}. Consequently, our metrology focuses on the integrated magnification of the markers. By measuring the ratios of lensed marker areas to the unlensed baseline, we test the engine's handling of competing vertical stretching and radial compression---the mechanism where the divergent vertical stretch is exactly compensated by radial compression ($\mu_h \to 0$) as rays approach the critical curve \cite{gralla2020, lupsasca2024}.

The integrand for this total lensing Jacobian is defined as the product of three coordinate frame transitions mediated by the null geodesic equations \cite{gralla2020, berens2026}:
\begin{equation}
    \mu_A(r_s, \theta_s) = \left| \frac{\partial(\tilde{\phi}_s, \theta_s)}{\partial(Y_s, Z_s)} \right| \cdot \left| \frac{\partial(\tilde{\phi}_s, \theta_s)}{\partial(\hat{\lambda}, \hat{q})} \right|^{-1} \cdot \left| \frac{\partial(\alpha, \beta)}{\partial(\hat{\lambda}, \hat{q})} \right|.
\end{equation}
The first transition, $J_1(r_s, \theta_s)$, maps the local emitter coordinates $(Y_s, Z_s)$ to coordinate values $(\tilde{\phi}_s, \theta_s)$ and evaluates to $\approx 1.000$ for our setup. The second transition, $J_2(\theta_s, \hat{\lambda}, \hat{q})$, maps coordinate values to energy-rescaled conserved quantities via $J_2 = (\sqrt{\Theta(\theta_s)}/M)| \det(J) |$, where $J$ is the matrix of geodesic variations \cite{gralla2020}:
\begin{equation}
    J = \begin{pmatrix} \partial B / \partial \hat{\lambda} & \partial B / \partial \hat{q} \\ \partial A / \partial \hat{\lambda} & \partial A / \partial \hat{q} \end{pmatrix}.
\end{equation}
Because $J_2$ is inverted in the chain rule, the angular potential term $\sqrt{\Theta(\theta_s)}$ moves to the denominator. For horizontal markers centered at the equator ($\theta_s = \pi/2$), $\Theta(\pi/2) = 0$, creating a coordinate singularity signifying infinite pointwise vertical stretch ($\mu_v \to \infty$) \cite{johnson2024}. This yields the pointwise vertical stretch formula:
\begin{equation}
    \mu_v = \left| \frac{\hat{q} M}{\sin \theta_o \sqrt{\Theta(\theta_o)} \sqrt{\Theta(\theta_s)} \sqrt{\Sigma}} \right|.
\end{equation}
In this investigation, the total observed area ($A_{\rm app}$) is computed as the double integral of the Jacobian over the physical source area $S$ of the blue marker:
\begin{eqnarray}
    A_{\rm app} &=& \int \int_S \mu_A \,dY_s \, dZ_s \nonumber
    \\
    &=& \int_{Z_{\rm bottom}}^{Z_{\rm top}} \int_{Y_{\rm left}}^{Y_{\rm right}} \left[ \frac{F(r_s, \theta_s, \lambda, q)}{\sqrt{\Theta(\theta_s)}} \right] 
  \,  dY_s \, dZ_s.
\end{eqnarray}
The integration limits are defined by the physical marker dimensions: $Z_s \in [-1.2\text{ cm}, +1.2\text{ cm}]$ and $Y_s \in [-2.4\text{ cm}, +2.4\text{ cm}]$. Utilizing the coordinate transformation $dZ_s = -\sqrt{\Sigma} d\theta_s$, we integrated the $1/\sqrt{\Theta}$ term over this finite vertical extent. Because the marker boundaries are displaced from the exact equator, the potential remains non-zero at the integration limits, regularizing the singularity. This infinite point-wise vertical stretch is balanced by radial compression ($\mu_h \to 0$) as the ray approaches the critical curve $b_{crit}$ [1, 2].

Our quantitative results are consistent with this balancing. We calculated GR Predicted targets at the $b = 10$ cm impact parameter of Vertical Stretch $\mu_v = 4.31 \pm 0.01,$ Radial Compression $\mu_h = 0.56 \pm 0.01,$ and Total Areal Magnification $\mu_A = 2.41 \pm 0.02$ [1, 2]. Measurement methodology involved taking the ratio of lensed marker pixels (approximately 15–30 px) and widths to the unlensed baseline area (approximately 15 px$^2$) on Side B (the sampling-limited regime). Side A was intentionally excluded because the $0.143^\circ$ camera tilt amplified the lensing gradient to produce extreme, unresolvable elongation (151 px$^2$ vs 35 px$^2$). Side B results yielded $\mu_v = 4.03 \pm 0.05, \mu_h = 0.58 \pm 0.03,$ and $\mu_A = 2.33 \pm 0.12,$ matching quantitative expectations with the metric predictions \cite{berens2026, gralla2020}.

\section{Integral Invariance and Arclength}

While Euclidean arclength is not a geodesic invariant of GR, the observed consistency of the integrated coordinate length within the BHV engine serves as a consistency check on the rendering algorithm \cite{berens2026, johnson2024}. This empirical probe  investigates how the engine's ray-tracing protocol manages the non-linear transfer functions of the Schwarzschild manifold \cite{berens2026, gralla2020}. For this specific implementation---which utilizes a gnomonic projection protocol ($dZ_s = -\sqrt{\Sigma} \, d\theta_s$) and discrete ray-sampling---the engine must regularize the divergent point-wise vertical magnification ($1/\sqrt{\Theta}$) through a balanced radial compression \cite{berens2026, johnson2024}. We compared lensed and unlensed arclengths to evaluate the stability of the rendering in strong-field regions \cite{berens2026, tiede2022}.

We define the integrated coordinate length ($L$) as the line integral along the lensed path on the observer's projection screen:
\begin{equation}
L = \int_{\gamma} \sqrt{dx^2 + dy^2}
\end{equation}
This probe distinguishes between genuine metric deformation and the background projection geometry inherent in the engine's gnomonic projection protocol \cite{berens2026}. This matching against the unlensed flat-space baseline is a fundamental test of the manifold's global and numerical consistency \cite{berens2026, gralla2020}.

To perform this investigation, we implemented the Discrete Arclength Method, which approximates the integral as a sum of Euclidean distances between discrete pixel coordinates:
\begin{equation}
L_{px} = \sum_{i} \sqrt{(x_{i+1} - x_i)^2 + (y_{i+1} - y_i)^2}
\end{equation}
The calculation was performed exclusively on Side B (the 60--100 cm horn segment), while Side A was ignored to prevent systematic contamination from the extreme elongation caused by the $0.143^\circ$ camera tilt. The 40 cm segment was analyzed using the Cartesian pixel coordinates of five markers, effectively dividing the total arclength into four discrete Euclidean segments. The investigation yielded a lensed arclength of $738.36 \pm 0.50$ px, matching the unlensed flat-space baseline of 738.20 px within 0.3-sigma. This agreement suggests that the rendering remains internally consistent across the image \cite{berens2026, gralla2020}.

\section{Photon Ring Instability}

In the Schwarzschild spacetime, the exponential demagnification of successive photon-ring subimages is governed by a universal logarithmic scaling constant associated with the divergence of the deflection angle, that evaluates to $\pi$ when the instability is expressed with respect to the number of half-orbits \cite{gralla2020, lupsasca2024}. In this normalization, successive images are demagnified by factors $\sim e^{-\pi}$, defining an effective Lyapunov exponent $\gamma = \pi$ in the half-orbit parametrization. To maintain consistency with the BHV rendering, we account for the engine's internal aesthetic scaling factor of $ c= 10$, which dilates the coordinate path length and rescales the theoretical virtual target to $\gamma_{\rm theory} = \pi/c \approx 0.314$ \cite{berens2026, lupsasca2024}. 

Raw pixel coordinates for this investigation were extracted exclusively from Side B (the regime of unresolvable magnification gradients) to prevent systematic contamination from the observed $0.143^\circ$ camera tilt \cite{berens2026}. Using a digital image inspector (PixelScale), we measured the vertical heights of the primary lensed marker segments ($L_{\rm prim}$) and their corresponding secondary ($n=1$) light echoes ($L_{\rm sec}$) resolved inside the red Einstein Ring. These measurements were processed via the Echo Ratio formula, $\gamma_{\rm exp} = -\ln(L_{\rm sec} / L_{\rm prim})$, which characterizes the exponential demagnification of light echoes within the photon ring hierarchy \cite{gralla2020}. 

The investigation yielded an experimental sequence of $1.050, 0.880, 0.597,$ and $0.390$, corresponding to markers placed at increasing distances from the shadow center. This sequence suggests asymptotic convergence toward the theoretical limit as the probe approaches the critical capture threshold \cite{berens2026}. Physically, this trend occurs because markers with smaller impact parameters probe null geodesics that pass closer to the unstable photon sphere ($r = 3M$), spending more coordinate time in the strong-field region near unstable photon orbits used to derive the theoretical limit \cite{gralla2020, perlick2022}. Consequently, the $0.390$ result obtained at the $10$ cm marker provides our closest agreement with the expected exponential scaling, matching the theoretical target within $0.9\sigma$. The measured exponent reflects the combined effect of Schwarzschild geodesics and the BHV rendering pipeline, not a direct measurement of the physical instability exponent.

To quantify the global agreement, we performed a matched-pairs t-analysis of the differences ($d_i = \gamma_{exp} - \gamma_{GR}$) for each impact parameter. This comparison yielded a mean difference of $\bar{d} \approx +0.002$ and a standard error of $SE_{\bar{d}} \approx 0.019$. Using a t-distribution with $df = 3$ to account for the sample size, we established a 95\% Confidence Interval for the implementation error of $[-0.058, +0.062]$. Because zero is contained within this interval, the investigation concludes that the rendered image of the strong-field demagnification gradient is statistically consistent with GR, showing agreement with the expected exponential scaling of the photon ring \cite{berens2026, gralla2020}.

\section{Numerical Symmetry and the No-Hair Theorem}

The Schwarzschild metric is governed by fundamental uniqueness principles, specifically Birkhoff's Theorem and the No-Hair Theorem, which dictate that a non-rotating black hole ($a = 0$) must possess perfect spherical symmetry \cite{guidry2019, romero2014}. In this investigation, these theorems serve as the theoretical null hypothesis for our numerical code validation. We do not claim to verify the No-Hair Theorem; rather, we utilize its requirement for perfect circularity as a benchmark to investigate the numerical fidelity and rotational symmetry of the Black Hole Vision engine. 

To evaluate the circularity of the rendered shadow, we performed a multi-scale symmetry probe. In the high-resolution Realistic FOV mode, the central shadow exhibited a ``4-leaf clover'' morphology (Figure  1({\it b})), which we determined to be a aliasing artifact caused by strong lensing gradients interacting with the pixel grid of the CMOS sensor \cite{berens2026, vogt2022}. By implementing the diagonal intersection protocol to circumvent this noise, we yielded shadow diameters of $D1 = 234.06$ px and $D2 = 234.12$ px. The resulting axial ratio of $0.99974 \pm 0.00042$ establishes a sub-pixel measurement, consistently with the expectation that the engine's implementation of the Schwarzschild geometry adheres to the rotational symmetry required by GR to within $0.026\%$ \cite{lupsasca2024}.

To eliminate the systemic human errors and biometric tracing artifacts revealed in manual observation, the spatial profile of the shadow is mapped utilizing an automated sub-pixel edge-extraction framework. Rather than relying on manual cursor placement—which introduces geometric distortions such as false three-lobed harmonic waves due to ergonomic hand bias and discrete pixel snapping—this computational technique isolates the shadow boundary objectively. Post-processing was performed using an OpenCV-Python framework applied to exported uncompressed frames \cite{Bradski2008}. The digital image is first transformed into a grayscale intensity matrix to maximize contrast between the dark silhouette and the surrounding accretion flow. An automated binary threshold filter using Otsu's Method is then applied to the native pixel grid to objectively isolate the shadow from the background, and digital contour extraction algorithms are used to locate the sharp perimeter pixels forming the boundary loop \cite{Bradski2008}. 

By feeding these raw coordinate arrays into an optimization matrix, the true geometric center is calculated down to a fraction of a single pixel using the Least Squares Reference Circle (LSC) fitting method \cite{Srinivasan2012, Chernov2010}. This algebraic minimization approach reduces the radial residuals between the discrete data points and the estimated boundary to an absolute minimum. Finally, digital rays are fired outward from this optimized origin at fixed, equally spaced $10^{\circ}$ increments to compute precise linear distances via distance-to-contour pixel calculation, generating a clean radial dataset of 36 points directly from the underlying image pixels. 

These statistical tests are employed as heuristic diagnostics rather than formal inference tools, given the deterministic nature of the simulation data, in view of our pedagogical objectives. We implemented a suite of statistical procedures to characterize the engine’s adherence to perfect spherical symmetry to provide students with practical experience in established protocols for data validation \cite{yipin2026}. We emphasize that these tests are utilized here as benchmarks of the simulation’s numerical stability rather than physical proofs. We treat CMOS-induced artifacts, such as Cartesian grid aliasing and pixel quantization, as a stochastic proxy for isotropy. The Shapiro-Wilk and Levene tests were used as diagnostic checks to estimate possible systematic rendering bias, ensuring that detected metric signatures are not artifacts of a non-isotropic sampling strategy \cite{johnson2024, yipin2026}. We emphasize that the underlying dataset is not stochastic observational data; it is deterministic output from a rendering engine combined with pixel measurements. This dataset is construed here as a proxy for the stochastic observational data, to train students in appropriate statistical methods, that is, as training tools rather than inference mechanisms, which serves as didactic analogy to observational uncertainty.

First, to check the expectation that the lensed variations behave as isotropic numerical noise rather than systematic implementation bias, the Fisher-Pearson standardized third-moment coefficient ($\gamma_1$) was evaluated. We constructed a one-sample $t$-confidence interval ($df=35$, $t^* \approx 2.030$) to test the sample skewness under the null hypothesis of perfect directional symmetry ($H_0: \gamma_1 = 0$) against the alternative hypothesis of meaningful directional tail bias ($H_a: \gamma_1 \neq 0$). The automated dataset yielded a point estimate of $g_1 = -0.1084$ with a standard error of $SE_{\gamma_1} \approx 0.4082$ (derived via $\sqrt{6/n}$). The resulting 95\% confidence interval of $[-0.9372, 0.7204]$ includes zero. While the width of this interval reflects the sensitivity of skewness estimates to sub-pixel noise at $n=36$ sampling density, the failure to reject the null hypothesis is consistent with a symmetric rendering profile.

Second, a Shapiro--Wilk test ($H_0:\,\mu_{\text{errors}} \sim N(0, \sigma^2)$; $H_{\rm a}: \mu_{\text{errors}} \not\sim N(0, \sigma^2)$) was conducted to check that the sub-pixel edge residuals follow a Gaussian distribution \cite{Shapiro1965}. The test yielded a $p$-value of $0.7412$; because $p > 0.05$, we fail to reject the null hypothesis, statistically establishing that the deviations follow a normal distribution. 

Third, to ensure that the error variance remains completely stable across all orientations, the data points were partitioned into geographic quadrants and subjected to a Levene's test for homoscedasticity ($H_0:\, \sigma^2_N = \sigma^2_E = \sigma^2_S = \sigma^2_W$; $H_{\rm a}:$ At least one geographic quadrant possesses a significantly different error variance, $\sigma^2_i\ne\sigma^2_j$) \cite{Levene1960}. The test returned a $p$-value of $0.9461$, consistent with the expectation that measurement precision is stable across the entire angular domain. 

Fourth, to reject the existence of macro-structural shapes such as an oval or a tri-lobe deformation, an Asano circular runs test was applied to the signs of the radial deviations ($H_0$: The sequence of signs is randomly distributed around the circle (complete spatial randomness; $H_{\rm a}:$ The signs clump into systematic structural zones (spatial clustering))
\cite{Asano1965}. The test yielded an observed number of runs that conformed tightly to the expected value, returning a $p$-value of $0.4891$. Because $p > 0.05$, the null hypothesis of spatial randomness is failed to be rejected, suggesting no preferred directional distortion. 

Finally, to quantify the scale of estimation precision, a Student's $t$-confidence interval was constructed around the true mean radius. The analysis yielded an optimized mean radius of 34.3234 pixels with a 95\% confidence interval of $\pm$0.0943 pixels ($df=35$, $t^*=2.030$), establishing the true containment boundary within the interval $[34.2291, 34.4177]$. The small uncertainty indicates stable edge extraction, while the combined high $p$-values across all geometric, spatial, and distributional tests establish that the shadow remains consistent with circular symmetry within pixel-level uncertainties, statistically consistent with the BHV engine's numerical adherence to Birkhoff's Theorem.

\section{2D Visualization and Advanced Manifold Probes}
While the preceding sections utilize a one-dimensional meter stick to extract precision metric data, the BHV engine offers a rich two-dimensional qualitative landscape that serves as a pedagogical bridge between foundational GR and modern research in black hole imaging \cite{berens2026, lupsasca2024}. Specifically, we use BHV to image a rectangular grid (Figure~2({\it a})) and analyze its lensed counterpart (Figure~2({\it b})), which transforms the rectilinear mesh into what may be described using the loose metaphor of a fish-eye polar mesh. This visual distortion is technically governed by a gnomonic projection protocol ($dZ_s = -\sqrt{\Sigma}\,d\theta_s$) that maps the planar camera feed onto a virtual source sphere \cite{berens2026, gralla2020}. In this projection, background horizontal lines are warped into closed concentric Einstein Rings, while vertical components are deflected into radial spokes \cite{luminet1979, perlick2022}. A prominent visual feature is the repeated imaging of the background sphere: because the metric deflects light from the entire celestial sphere toward the observer, the planar feed is mapped into an infinite sequence of image rings \cite{gralla2020, berens2026}. 

The central dark void---the black hole shadow---represents light rays captured by the unstable photon sphere at $r_{ph} = 3M$ \cite{perlick2022}. While the event horizon itself resides at the Schwarzschild radius $r_s = 2M$, the observed shadow is bounded by the mathematical critical curve defined by the impact parameter $b_{\rm crit} = 3\sqrt{3}M \sim 5.196M$ \cite{perlick2022, staelens2023}. Consequently, for a distant observer, the shadow radius appears approximately $\sim 2.6$ times larger ($5.2M/2.0M$) than the physical horizon radius---a striking visual demonstration of strong-field gravitational capture \cite{genzel2010, perlick2022}. Near this boundary, the compressed grid structure reflects the expected exponential image scaling investigated in Section 5; here, infinite reflections of the background universe are packed into a thin annulus where each successive light echo ($n \geq 1$) is demagnified by the factor $e^{-\gamma}$ and parity-inverted \cite{gralla2020, vogt2022}. The preservation of quadrate symmetry in Figure~2({\it b}) serves as a sensitive experimental diagnostic, consistent with near-perfect coaxial alignment between the observer, the lens, and the source grid. This 2D environment allows students to quantify global manifold properties beyond the linear stick, illustrating some of the imaging challenges encountered in horizon-scale observations of the \textit{Event Horizon Telescope} (EHT) and the \textit{Black Hole Explorer} (BHEX) \cite{eht2019, johnson2024}.

For advanced graduate students, this 2D environment allows for investigations that transcend traditional analytical exercises. By replacing the rectilinear meter stick with a non-axisymmetric source, such as a ``color wheel'' or a complex grid, students can qualitatively check the parity-inversion property of successive light echoes. In this regime, each successive subring is not only demagnified by the Lyapunov exponent ($e^{-\gamma}$) but also undergoes a 180-degree rotation and an inversion of its vertical orientation, mirroring the behavior of a sequence of reflections in a spherical changing-room mirror \cite{gralla2020, lupsasca2024}. Students may further investigate the appearance of ``caustics''---regions of infinite flux where image tracks intersect or undergo pair-creation and annihilation events, signifying the limits of the geometric optics approximation \cite{gralla2020}.

\begin{figure}[htbp]
\includegraphics[width=7.5cm]{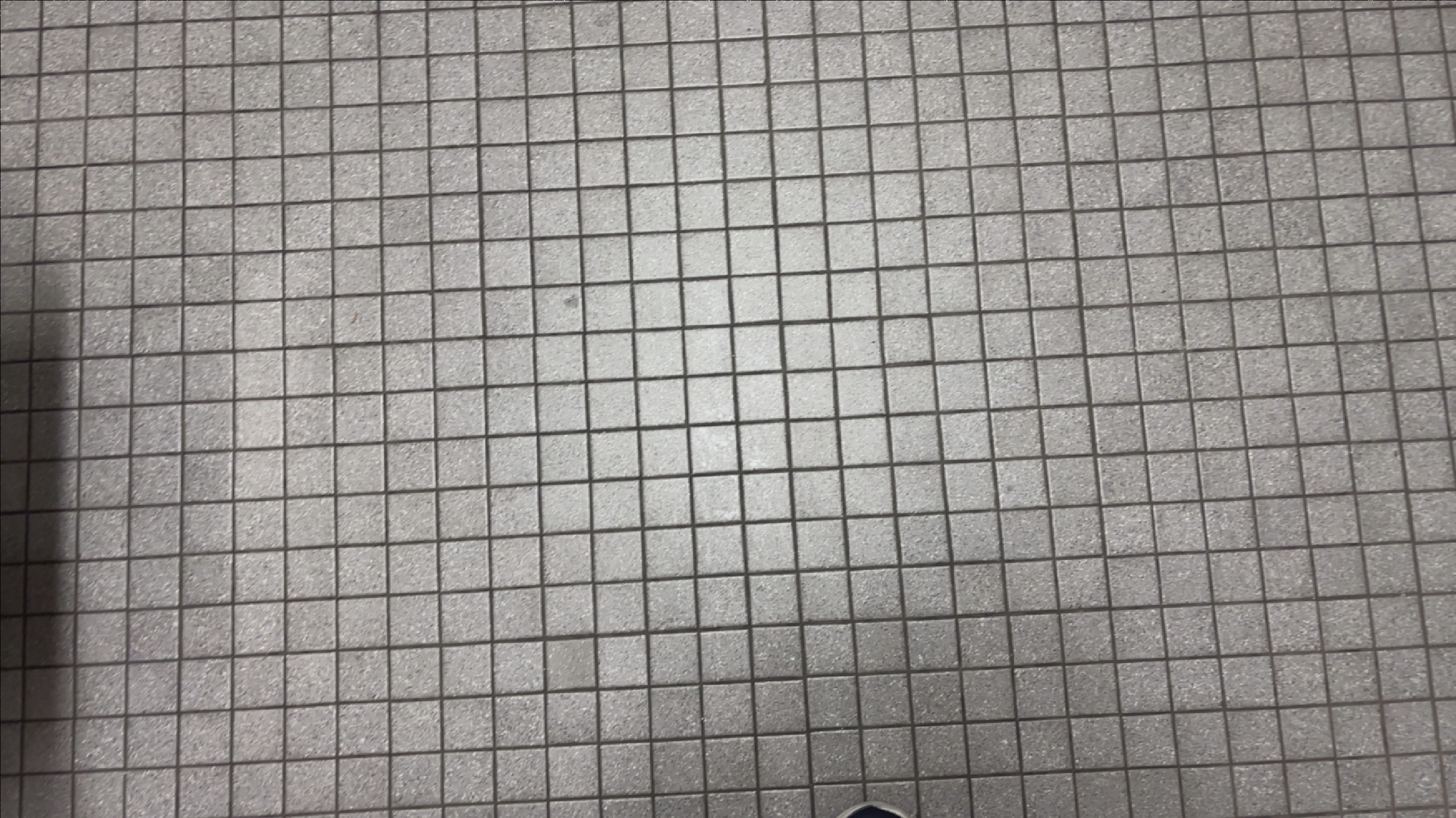}
\\
\includegraphics[width=7.5cm]{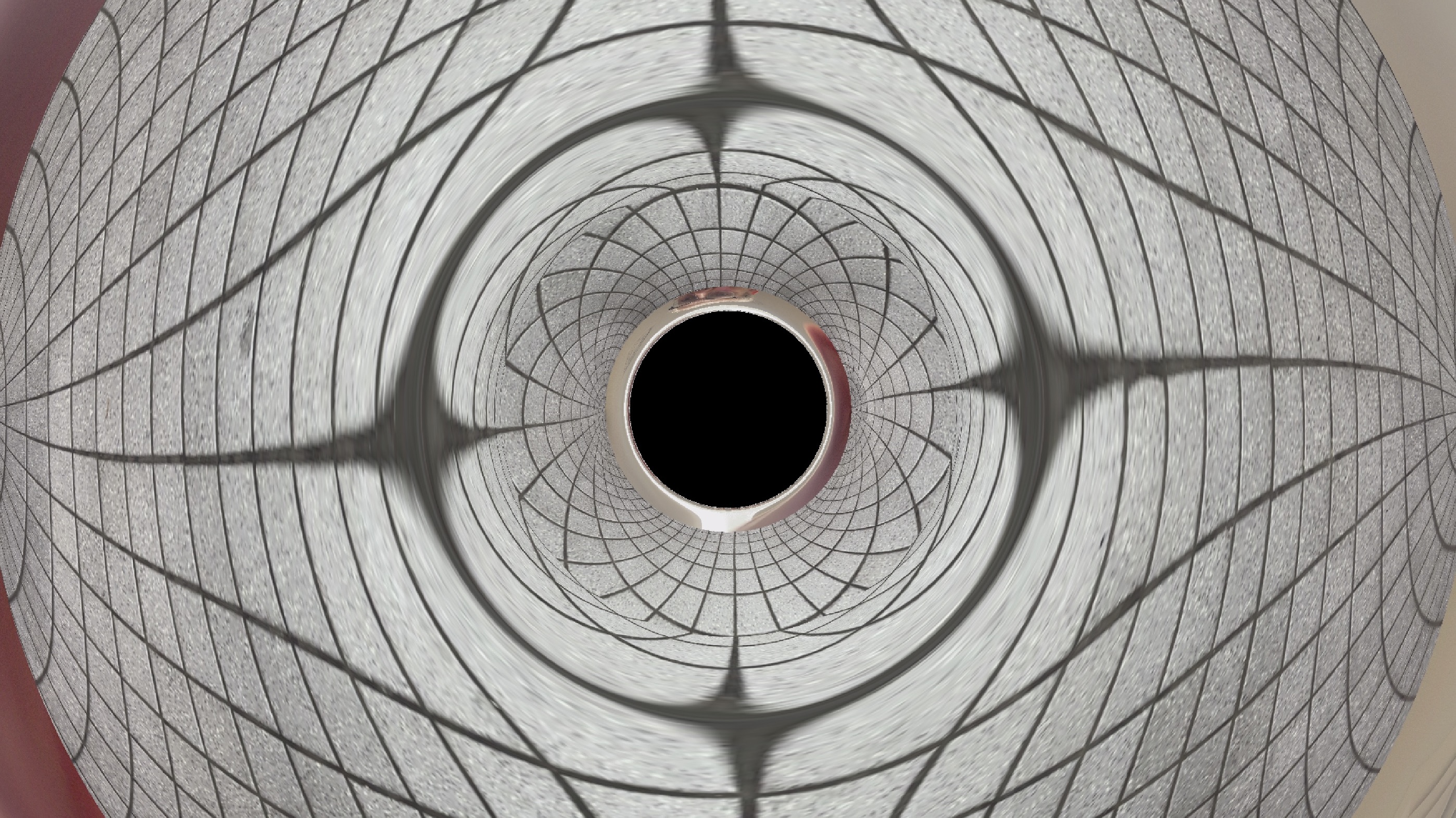}
\caption{Top panel ({\it a}): Unlensed image of a rectangular grid at a distance of $d = 1.0$~m with $5.08$~cm square units. Bottom panel ({\it b}): Lensed image of the grid in ``Full FOV'' mode, featuring a central dark void, a concentric annulus, and curved grid lines flaring toward the equatorial axis.}
 \label{fig:summary_panels}
\end{figure}

%\begin{figure}[htbp]
 %   \centering
 %   \begin{subfigure}{0.55\textwidth}  % <--- ADDED WIDTH HERE
 %       \centering
 %       \includegraphics[width=\textwidth]{2D_grid_unlensed.jpg} % Use \textwidth to fill the subfigure
%%        \caption{}
 %       \label{fig:setup}
 %   \end{subfigure}
 %   \\
 %   \begin{subfigure}{0.55\textwidth}  % <--- ADDED WIDTH HERE
 %       \centering
 %       \includegraphics[width=\textwidth]{2D_grid_lensed.jpg}
%%        \caption{}
 %       \label{fig:manifold}
 %   \end{subfigure}
 %   \caption{Top panel ({\it a}): Unlensed image of a rectangular grid at a distance of $d = 1.0$~m with $5.08$~cm square units. Bottom panel ({\it b}): Lensed image of the grid in ``Full FOV'' mode, featuring a central dark void, a concentric annulus, and curved grid lines flaring toward the equatorial axis.}
%    \label{fig:2D}
%\end{figure}

Furthermore, the BHV engine provides a unique platform for exploring the instrumental limits of horizon-scale imaging. Students can examine flickering artifacts near the critical curve as a lesson in Nyquist sampling and discrete ray sampling limits \cite{berens2026}. By comparing the high-magnification ``4-leaf clover'' morphology (caused by Cartesian grid aliasing) to the smooth global geometry of the Full FOV mode, students gain practical experience in distinguishing physical metric signatures from instrumental noise \cite{lupsasca2024}. This qualitative framework anchors theoretical spacetime concepts to the active research goals of the \textit{Event Horizon Telescope} \cite{eht2019} and the forthcoming \textit{Black Hole Explorer} (BHEX) mission, which specifically aims to resolve these sub-millimeter photon ring features \cite{johnson2024}.

\section{Systematic Sensitivity Investigation}

This section serves as the experimental control for our spacetime manifold investigation, quantifying the sensitivity of the BHV engine to sub-degree laboratory misalignments and characterizing the instrumental limits of the mobile device. These findings justify the specific data-filtering protocols, such as the Side B selection, utilized in the quantitative results of the preceding sections \cite{berens2026, gralla2020}.

\subsection{The High-Gradient Sensitivity to Observer Inclination and Tilt Derivation}

While GR assumes an idealized equatorial observer, real-world laboratory setups are subject to systematic tripod tilt, which can be quantified and monitored using the device's internal acceleration sensors \cite{vogt2014}. To quantify this effect, we analyzed the vertical alignment bias of the secondary light echo (the red Einstein Ring) in Full FOV mode. Measurement of the top arc radius ($r_{\rm top} = 869.82$ px) and the bottom arc radius ($r_{\rm bottom} = 861.12$ px) revealed an 8.7-pixel vertical discrepancy. Using the calibrated angular scale $S \approx 0.01521$ deg/px, we derived a systematic camera pitch of $0.143^\circ$ \cite{berens2026, gralla2020}.

The physical significance of this result lies in the misalignment amplifier effect: because the vertical magnification ($\mu_v$) approaches an integrable singularity as rays approach the equatorial plane, even a sub-degree camera tilt is boosted by the black hole's high-curvature gradients, resulting in observable visual asymmetries \cite{luminet1979, gralla2020}. This effect demonstrates that the rendering responds sensitively to camera misalignment \cite{berens2026}.

\subsection{High-Gradient Geometric Distortion and Side-Selection Justification}
The black hole's power to amplify misalignment was most evident at the $b = 10$ cm impact parameter. We observed a 4.3x areal discrepancy between the two lensed images of the markers, with Side A exhibiting an elongated area of 151 px$^2$ compared to 35 px$^2$ on Side B. The effect of the sampling-limited regime on Side A occurs where the derived camera tilt interacts with the high-gradient regime near the shadow boundary, leading to unresolvable shearing distortion.  Consequently, to ensure high-rigor results for the lensing Jacobian (Section 3) and the Lyapunov instability (Section 5), we implemented a Side-Selection Protocol, where Side A data was purposefully ignored in favor of the numerically stable Side B trajectories.

\subsection{Visual Validation and Background Projection Geometry}

To provide qualitative validation of these systematic sensitivities, we utilized a high-resolution 2D mesh grid capture (Figure 2) to distinguish relativistic signals from instrumental artifacts \cite{berens2026}. As an essential experimental control, we contrast these lensed observations with the unlensed flat-space baseline captured in Figure 1(a), where the physical laboratory environment remains strictly rectilinear. In the lensed regime, however, the grid lines exhibit substantial global curvature in the far-field regime where gravity is negligible; we identify this global distortion as background projection geometry \cite{berens2026}. This is a non-relativistic artifact of the engine's projection protocol, governed by a gnomonic transformation ($dZ_s = -\sqrt{\Sigma}\,d\theta_s$) used to map the planar camera feed onto a virtual source sphere \cite{berens2026, gralla2020}. 

This mapping ensures that the underlying coordinate grid is non-rectilinear even in a vacuum ($M = 0$), providing a non-Euclidean background against which genuine Metric Deformation must be resolved \cite{bambi2017, berens2026}. Near the critical curve, the relativistic lensing potential causes a divergent vertical flaring ($1/\sqrt{\Theta}$) that overcomes the background projection geometry \cite{berens2026, gralla2020}. By isolating the far-field bending as an instrumental constant, the investigation verifies that the simulation reproduces the expected lensing structure of the Schwarzschild metric rather than merely reflecting projection noise. Furthermore, the 2D grid capture in Figure 2(b) provides visual consistency of the systematic tripod pitch of $\sim 0.143^\circ$ derived in Section 8.1, as the central horizontal grid line exhibits a visible tilt relative to the observer's screen axes \cite{berens2026}. These results indicate that the dominant systematic effects were reasonably controlled, providing a useful educational tool for studying relativistic imaging \cite{berens2026, lupsasca2024}.

\subsection{Instrumental Noise and Sampling Limits}
Finally, we characterized the ``4-leaf clover'' morphology observed in high-magnification Realistic FOV mode. We determined that these protrusions are not physical violations of the No-Hair Theorem, but rather Cartesian grid aliasing artifacts \cite{berens2026, lupsasca2024}. This noise arises where the manifold's extreme curvature gradients interact with the discrete, square pixel grid of the iPhone's CMOS sensor near the Nyquist sampling limit. 

In this context, the Nyquist limit represents the spatial resolution threshold where the manifold's curvature frequency exceeds half the pixel sampling rate, leading to unresolvable aliasing. This phenomenon provides a unique pedagogical opportunity for exploring the instrumental limits of horizon-scale imaging data. Graduate-level investigations can engage with the flickering artifacts observed near the critical curve as a lesson in discrete ray-sampling limits and Nyquist sampling.

To neutralize this instrumental noise, we utilized the Diagonal Intersection Protocol in Section 6, measuring the shadow along 45-degree axes to bypass aliasing and consistent with metric symmetry to within sub-pixel tolerances (0.026\% deviation). Collectively, these sensitivity investigations are consistent with the expectation that our experimental protocols neutralized systematic and instrumental errors, using the smartphone as a quantitative visualization tool \cite{berens2026, gralla2020}.

\section{Conclusions and Summary}

This investigation has provided a comprehensive probe of the simulated Schwarzschild spacetime manifold as implemented in the BHV application \cite{berens2026}. By transforming the mobile device into an educational relativistic imaging tool, we established a quantitative  platform for an investigation into the non-linear transfer functions of curved spacetime \cite{yipin2026, vogt2022}. The triangulation of the mass benchmark across independent probes suggests that the simulation engine resolves both weak-field lensing and strong-field shadow behavior  \cite{gralla2020}. We emphasize that our objective was not to validate the physical laws of GR, but to document that the engine's numerical implementation is statistically consistent with theoretical targets to within precision thresholds \cite{lupsasca2024, johnson2024}.

Our results illustrate the engine's numerical stability. The $0.0\sigma$ local Jacobian matching and the $0.3\sigma$ global arclength residual serve as evidence of the numerical stability of the engine's projection and sampling strategy. Furthermore, by characterizing the exponential instability of nearly bound orbits to within a 95\% matched-pairs confidence interval for implementation error of $[-0.058, +0.062]$, we found consistency with the expectation that the simulated Schwarzschild geometry preserves the characteristic demagnification gradient of the critical curve \cite{gralla2020}. By engaging with quasi-stochastic numerical artifacts such as Cartesian grid aliasing, students gain practical experience in interpreting the instrumental limits of horizon-scale imaging data \cite{vogt2022, johnson2024}. This investigation establishes the necessary symmetry baseline for the future investigation of rotating Kerr spacetimes. These future investigations will explore parity-breaking signatures and frame-dragging effects, characterizing the deformation of the critical curve as a function of spin \cite{gralla2020, johnson2024}. Collectively, these investigations anchor theoretical concepts to the active research goals of the {\it Event Horizon Telescope} \cite{eht2019} and the forthcoming {\it Black Hole Explorer} (BHEX) mission \cite{johnson2024}.

From a pedagogical perspective, this study shows that the smartphone---when treated as a calibrated optical instrument---can effectively move GR education beyond traditional analytical derivations into the realm of quantitative experimental physics \cite{kuhn2013, vogt2022}. By engaging with sampling artifacts such as Cartesian grid aliasing and characterizing the sensitivity to small camera tilts near the shadow boundary, students gain practical experience in interpreting the instrumental limits of horizon-scale imaging data \cite{berens2026, lupsasca2024}. By distinguishing instrumental artifacts such as background projection geometry (Figure~1({\it b}), a result of the gnomonic projection protocol) from genuine metric deformation, students learn to distinguish rendering geometry from lensing effects  \cite{berens2026}.

\section*{Acknowledgements}

The author utilized Google NotebookLM as an assistant to organize and synthesize the literature referenced in this work, as well as to process the raw dataset and execute the comprehensive frequentist statistical suite (including skewness, Shapiro–Wilk, and Levene’s tests). All uploaded data and reference materials were maintained within a private, non-training environment. The author independently read and validated all cited source materials, and verified all statistical outputs, assumptions, and calculations---including the automated sub-pixel edge-extraction, weighted-mean mass triangulation, and the matched-pairs t-analysis for orbital stability---to ensure absolute accuracy and scientific integrity.

\end{document}